# Tying Process Model Quality to the Modeling Process: The Impact of Structuring, Movement, and Speed[1]

Jan Claes[1], Irene Vanderfeesten[2], Hajo A. Reijers[2], Jakob Pinggera[3],
Matthias Weidlich[4], Stefan Zugal[3], Dirk Fahland[2], Barbara Weber[3], Jan Mendling[5]
and Geert Poels[1]

[1] Ghent University, Belgium
{jan.claes, geert.poels}@ugent.be
[2] Eindhoven University of Technology, The Netherlands
{i.t.p.vanderfeesten, h.a.reijers, d.fahland}@tue.nl
[3] University of Innsbruck, Austria
{jakob.pinggera, stefan.zugal, barbara.weber}@uibk.ac.at
[4] Technion - Israel Institute of Technology, Israel
weidlich@tx.technion.ac.il
[5] Wirtschaftsuniversität Wien, Austria
jan.mendling@wu.ac.at

**Abstract.** In an investigation into the *process of process modeling*, we examined how modeling behavior relates to the quality of the process model that emerges from that. Specifically, we considered whether (i) a modeler's structured modeling style, (ii) the frequency of moving existing objects over the modeling canvas, and (iii) the overall modeling speed is in any way connected to the ease with which the resulting process model can be understood. In this paper, we describe the exploratory study to build these three conjectures, clarify the experimental set-up and infrastructure that was used to collect data, and explain the used metrics for the various concepts to test the conjectures empirically. We discuss various implications for research and practice from the conjectures, all of which were confirmed by the experiment.

## 1 Introduction

Business process modeling is utilized at an increasing scale in various companies. The fact that modeling initiatives in multinational companies have to rely on the support of dozens of modelers requires a thorough understanding of the factors that impact modeling quality [1–3]. One of the central challenges in this area is to provide modelers with efficient and effective training such that they are enabled to produce high-quality process models. There is clearly a need to offer operational guidance on how models of high quality are to be created [4, 5].

---

[1] The final publication is available at Springer via http://dx.doi.org/10.1007/978-3-642-32885-5_3

Recent research has investigated several factors and their influence on different measures of process model quality [6, 7]. In essence, this stream of research identifies both process model *complexity* and the reader's modeling *competence* as the major factors among these. While these insights are in themselves valuable, they offer few insights into how we can help process modelers to create better models right from the start. In order to give specific hints to the modeler, we have to shed light on how good process models are typically created, and in which way this creation process differs from drawing process models of lower quality.

In this paper, we look deeper into the modeling process in its relation to the creation of a high-quality process model. The research question we deal with, is whether it is possible to identify certain aspects of modeling style and model creation that relate to good modeling results. Our approach has been to leverage the Cheetah Experimental Platform [8], which allows for tracing the creation of process models on a detailed level. This permitted us to quantify the process of process modeling with respect to three different aspects. We also determined an objective measure for the quality of the resulting process models, putting the focus on the ease with which such models can be read. Based on an experiment with 103 graduate students following a process modeling course, we were able to demonstrate a strong statistical connection between three aspects of the modeling process on the one hand with our notion of model quality on the other. These findings have strong implications, as they pave the way for explicating and teaching successful modeling patterns.

The structure of the paper is as follows. Section 2 discusses cognitive concepts that are relevant for investigating the process of process modeling. In addition, we describe how the capabilities of the Cheetah Experimental Platform are conducive to document the process of process modeling in detail. Section 3 presents our research design. We explain how we developed three conjectures about process-related factors that result in better process models. Each of these three factors as well as the notion for process model quality is operationalized, such that the conjectures can be experimentally tested. Section 4 reports on the conduct and results of our experiment. We discuss the results and reflect upon the threats to their validity. The paper closes with conclusions and an outlook on future research.

## 2　　Background on the Process of Process Modeling

In this section, we revisit findings on process model quality and the process of process modeling. Section 2.1 summarizes prior research in this area, after which Section 2.2 discusses how the process of process modeling can be analyzed.

### 2.1　　The Process of Process Modeling and Process Model Quality

There is a wide body of literature that centers on the quality of process models, ranging from high-level, comprehensive quality frameworks (e.g., [3, 4, 9]) to a variety of metrics that pin down the quality notion in specific ways (e.g., [2, 10, 11]). Mostly, the process model is considered in these papers as a given, complete, and finished artifact. Recently, approaches are emerging that aim to connect the way that a process model has come into being with the properties of the ensuing model. In this context, various authors refer to the actual construction of a process model as *the process of process modeling* [8, 12, 13].

In general, modeling is often characterized as an iterative and highly flexible process [14, 15], dependent on the individual modeler and the modeling task at hand [16]. A central element in the further understanding of the process of process modeling is the identification of the recurring activities or common phases that comprise this process. Inspired by views on problem solving, Soffer et al. [13] distinguish between the phase in which a modeler forms a mental model of the domain and the phase in which the modeler maps the mental model to modeling constructs. The work presented in [16] is in line with this view by its explicit recognition of a *comprehension* phase and a *modeling* phase, yet extends it by the additional recognition of a *reconciliation* phase. During the latter phase, modelers may reorganize the process model at hand (e.g., rename activities) and utilize the process model's secondary notation (e.g., layout). While modeling and comprehension phases generally alternate, they may be interspersed with reconciliation actions [16]. In the same work, a so-called *modeling phase diagram* is introduced that can be used to categorize a modeler's actions using these phases.

At this point, several preliminary insights exist that relate the modeling process with the modeling outcome, i.e., the business process model. First of all, the structure of the informal specification that is used as the basis for a process modeling effort seems to be of influence on the accuracy of the ensuing process model [17]. The reason may be that pre-structuring such a specification lowers the mental effort for modelers, resulting in a process model that better reflects the actual domain. Another insight is that the specific reasoning tools that are at the disposal to the modeler, e.g., workflow patterns vs. behavioral patterns, seem to affect the mental model that the modeler creates of a domain and, in this way, influence the semantic quality of the process model [13]. Finally, in [18] it is empirically shown that providing modelers in a distributed setting with specific model building blocks will minimize model quality issues such as variations in terminology and abstraction that individual modelers use.

The work that is presented in this paper must be seen as an attempt to extend the list of factors that can be connected to the quality of a process model, in the spirit of [13, 17, 18]. Another similarity with these works is that an empirical angle is taken to investigate conjectures about the influence of attributes of the modeling process.

### 2.2 Tracing the Process of Process Modeling with Cheetah Experimental Platform

The process of process modeling can be analyzed by recording editor operations as a sequence of modeling events. In this paper, we rely on Cheetah Experimental

Platform[2]. This platform has been specifically designed for investigating the process of process modeling in a systematic manner [8]. In particular, the platform instruments a basic process modeling editor to record each user's interactions together with the corresponding time stamp in an event log, describing the creation of the process model step by step.

When modeling with Cheetah Experimental Platform, the platform records the sequence of adding nodes, i.e., activities, gateways and events, and edges to the process model, naming or renaming activities, and adding conditions to edges. In addition, modelers can influence the process model's secondary notation, e.g., by laying out the process model using move operations for nodes or by utilizing bend points to influence the routing of edges (see Table 1 for an overview of all recorded operations). By capturing all of the described interactions with the modeling tool, we are able to replay a recorded modeling process at any point in time without interfering with the modeler or her problem solving efforts. This allows for observing how the process model unfolds on the modeling canvas. We refer to [16] for technical details.

**Table 1.** Recorded events in Cheetah Experimental Platform and their classification

| **Create** | **Move** | **Delete** |
|---|---|---|
| CREATE_START_EVENT | MOVE_START_EVENT | DELETE_START_EVENT |
| CREATE_END_EVENT | MOVE_END_EVENT | DELETE_END_EVENT |
| CREATE_ACTIVITY | MOVE_ACTIVITY | DELETE_ACTIVITY |
| CREATE_XOR | MOVE_XOR | DELETE_XOR |
| CREATE_AND | MOVE_AND | DELETE_AND |
| CREATE_EDGE | MOVE_EDGE_LABEL | DELETE_EDGE |
| RECONNECT_EDGE (**) | CREATE_EDGE_BENDPOINT (*) | RECONNECT_EDGE (**) |
|  | MOVE_EDGE_BENDPOINT (*) |  |
|  | DELETE_EDGE_BENDBPOINT (*) |  |

Other : NAME_ACTIVITY, RENAME_ACITIVTY, NAME_EDGE, RENAME_EDGE

(*) create, move and delete edge bendpoint were considered as actions to move an edge

(**) reconnect edge was considered as deleting and creating an edge

## 3    Foundations of the Experimental Design

In this section we present the foundations of our experimental research design. Section 3.1 summarizes three conjectures that we derived from exploratory modeling sessions. Section 3.2 provides operational definitions for objectively measuring the process of process modeling. Section 3.3 builds an operational definition of quality for a resulting process model, which is suitable for our experimental setting.

---

[2] For download and information we refer to http://www.cheetahplatform.org.

### 3.1     Conjectures from Exploratory Modeling Sessions

To derive insights in the modeling process, we performed three small-scale experiments that involved 40 modelers in total. These were conducted at sites of the participating researchers throughout 2010. In these experiments modelers were asked to draw a process model on the basis of a given informal description, which was the same at all sites. We analyzed the results of these experiments by visualizing the recorded data in charts and by replaying individual modeling cases. To this end, we designed a visualization of the process of process modeling in terms of a PPMChart (Process of Process Modeling Chart)[3]. Fig. 1 is an example of such a chart. The horizontal axis represents a time interval of one hour. Vertically, each line represents one object of the model as it was present during modeling. Each dot represents one action performed on the object; the color of the dot represents the type of action: create, move, delete or (re)name. The objects are vertically sorted by the time of the first action; the first action performed on each model object is its creation. The dots are aligned to the right such that the last action performed by the modeler is shown to occur at the end of the one hour interval. In the example in Fig. 1, we observe a short process (about 17 min) where most of the model objects were moved after creation (second dot on many lines). Furthermore, we see that the modeler has worked in 'blocks', i.e. two activities were created followed by gateways and edges. Fig. 2 shows the clear and well-structured process model resulting from the creation process.

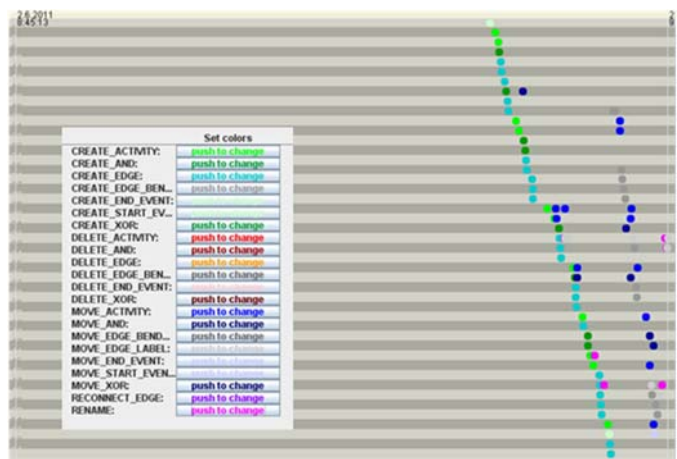

**Fig. 1.** Visualization of the operations in the creation of one model by one modeler.[4]

---

[3] We used the Dotted Chart Analysis plug-in of the process mining tool ProM for visualizing the PPMChart.
[4] High resolution graphs are available from http://bpm.q-e.at/paper/ModelQuality.

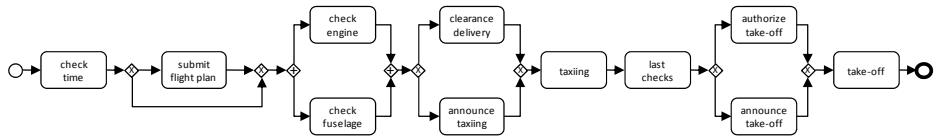

**Fig. 2.** Process model as result of the modeling process in Fig. 1

The interesting point of our exploratory session was the variation that we could observe in the PPMCharts. Fig. 3 shows different examples: Fig. 3a shows a process where objects were barely touched after creation, while Fig. 3b depicts a process with more actions, but mostly not long after the creation of the touched object. Fig. 3c shows a process where move actions occurred after creation of *all* objects. Fig. 3d visualizes a process with a rather chaotic actions pattern. Note that each PPMChart in Fig. 3 visualizes the creation of a process model based on the *same* textual description. It can clearly be observed, therefore, that some modelers create more elements, take more time to create their model, or move around objects on the canvas more frequently than other modelers.

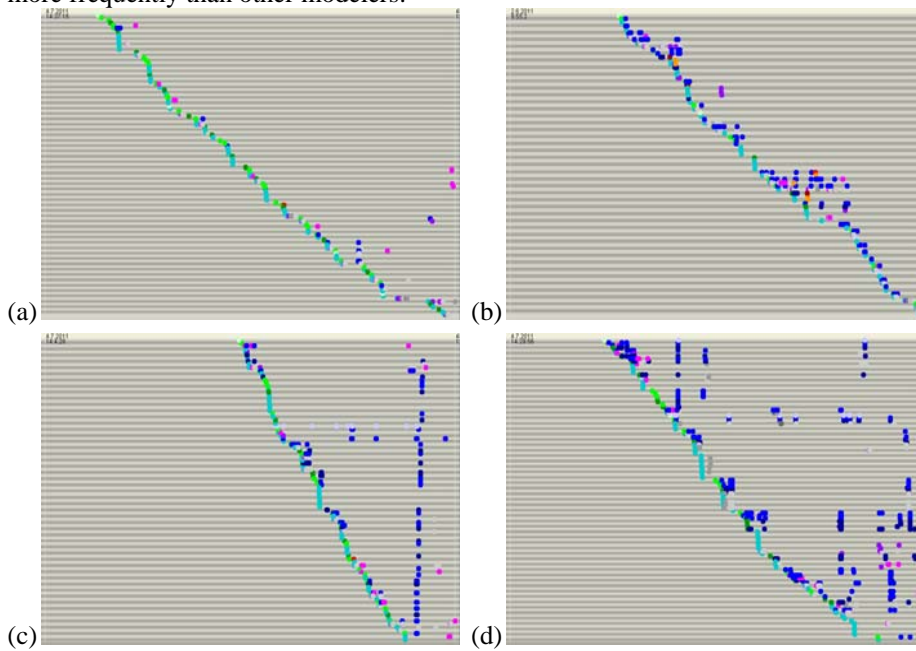

**Fig. 3.** More examples of the visualization of the operations in the process of process modeling.

The utilization of PPMCharts helped us to identify patterns of modeling and connections between the process of process modeling and the quality of the resulting process models. More specifically, we found three conjectures:

**Conjecture 1**: *Structured modeling is positively related with the understandability of the resulting model.*

The conjecture is related to the limited amount of items that humans can hold in their *working memory* [19]. Cognitive Load Theory suggests that problems arise when one's working memory is overloaded [20]. We therefore surmise that working on the complete model at once will make overloading of the working memory more likely, as compared to working on calculable pieces of the model, one at a time. Conjecture 1 defines this style of working as *structured modeling*. In other words, we assume that focusing on a specific, bounded part of the model (e.g., a block as apparent in the modeling process in Fig. 1) and finishing it before starting to work on another such part will help to reduce one's cognitive load. Hence, this style will result in better models.

**Conjecture 2**: *A high number of move operations is negatively related to the understandability of the resulting model.*

While studying the results of our exploratory experiments, we observed a notable difference in the structure of the modeling process across modelers. The data of the sessions suggest that modelers who frequently move model elements seem to have no clear idea in mind of how the process is supposed to be modeled. They will therefore potentially make more mistakes, which results models of lower quality.

**Conjecture 3:** *Slow modeling is negatively related to the understandability of the resulting model.*

Finally, we noticed a difference in the modeling speed of different modelers (i.e., in terms of the total time between the first and last recorded modeling actions). Presumably, modelers who are in doubt about the structure of the process or about the way to capture it, will spend more time thinking about the process, trying out different strategies to organize and re-organize the model. This will ultimately take more time to finalize the process model. We presume that the more time it takes the modeler to create the model, the lower the quality of the resulting model will be. Such an effect would be congruent with the result that faster programmers tend to deliver code with fewer defects than median or below-median performing programmers [21].

### 3.2 Operational Measurement of Process-based Factors

The challenge arising for these conjectures relates to their operational definition. For **Conjecture 1**, we need to provide an operational definition for a structured style of modeling based on the notion of blocks. In this context, a *block* consists of all involved model elements in two, or more, parallel or optional paths in the model. Mostly, this will concern a structure that consists of one split gateway, some successive activities, and one join gateway to complete it. We consider the modeling process to be *structured* if the modeler is not working on more than one block at the same time. The degree of structured modeling is determined based on the replay of the modeling process as visually assessed by an expert. This assessment provides the values of two metrics for structured modeling.

*MaxSimulBlock* is the maximum number of blocks that were simultaneously in construction. A block was considered in construction from the time the first element was created until the time the last element was created. If a block was changed afterwards (e.g., deleting and creating an activity), it had no effect on this metric.

*PercNumBlockAsAWhole* is the number of blocks that were made as a whole in relation to the total number of blocks. A block was considered to be made as a whole if no other elements (except for edges) were created between the creation of the first and last created element of the block.

We observed many modelers positioning activities and gateways in a block structure while adding the edges much later. For this reason, we did not consider the edges to be part of the block when calculating these metrics. As we are interested in the timing of the *creation* of elements in a block, we did not consider changes after the original creation of a block. Therefore, only those elements that were present at the initial completion of a block (this is the point in time when its last element is added) were considered to be part of the block.

For **Conjecture 2**, we consider how many elements were moved and how many moves were performed on these elements. This was calculated by a program that determined which of the recorded actions are move actions according to the list presented in Table 1. We define the following two metrics.

*AvgMoveOnMovedElements* is the average amount of move operations on elements with at least one move operation.

*PercNumElementsWithMoves* is the number of elements with move operations in relation to the total number of elements.

For **Conjecture 3**, we also wrote a small program to calculate the time spent until the model was finished. As we observed many modelers moving lots of elements around after finishing the creation of all elements, we distinguish the time between first and last action and between first and last *create* action.

*TotTime* is the total time between the first and last recorded action of the modeling process.

*TotCreateTime* is the total time between the first and last recorded *create* action of the modeling process.

### 3.3  Operational Measurement of Process Model Quality

There is a wide body of literature available on quality measures for process models. In this paper process model quality is defined as the ease with which the process model can be understood. In order to objectify this notion (and automate its assessment) we consider it from the structural correctness point of view; not from the semantical point of view. Prior research has defined an extensive amount of formal, structural correctness criteria for process models [22]. In the context of our experiments, we utilized BPMN as a modeling language. The problem with existing criteria, such as soundness, is that they are not directly applicable to BPMN models because BPMN does not enforce a WF-net structure [23]. Therefore, we consider a relaxed notion of quality, namely that the resulting process model should be *perspicuous*[5]. We operationalize the definition of a perspicuous model as *"a model that is unambiguously interpretable and can be made sound with only small adaptations based on minimal assumptions on the modeler's intentions with the model"*.

---

[5] See Merriam-Webster at http://www.webster.com/dictionary/perspicuous.

To make our notion of *model quality* robust against the familiarity of a modeler with notational conventions, we translate each model to a syntactically correct BPMN model whenever the model structure strongly hints at the modeler's intentions. The resulting BPMN model is then transformed into a WF-net according to the mapping defined in [24]. For such a WF-net, we checked soundness using LoLA [25]. A BPMN model is classified as being *perspicuous* if the respective WF-net is sound; otherwise, it is classified as *non-perspicuous*. In the remainder of this section, we describe the transformation to derive a syntactically correct BPMN model that can be transformed into a WF-net based on structural characteristics. The transformation is inspired by the preprocessing discussed in [24] and applied in the presented order[6].

**Handling of start and end events.** Many modeling languages do not have specific symbols for the start or end of the process (e.g., Petri-nets and EPCs). Modelers who are not aware of these specific events in BPMN may, therefore, forget to include them in their model. In line with the BPMN specification, we normalize such models:

- *Transform a process that does not have a start or an end event into a process that does, by preceding each task without incoming flows by a start event and succeeding each task without outgoing flows by an end event.* [24]

Further, some modeling languages allow for several starting points in the model (e.g., EPC, BPMN), cf. [26]. Also, it is allowed or even required that each end point in the process model is indicated separately (e.g., EPCs, COSA, BPMN). Modelers may be familiar with this explicit modeling of each start or end point, so that a WF-net structure is obtained by the following transformations:

- *Transform a process that has multiple start (end) events by replacing all start (end) events with only one start (end) event succeeded (preceded) by an XOR-split (XOR-join) gateway, and connect this gateway to each activity that was preceded (followed) by one of the original start (end) events.* [24]
- *If we determine only one origin for the multiple flows, i.e., all starting (ending) paths join in (originate from) the same gateway, we use the sign (i.e., AND or XOR) of this gateway.*

Note that the latter rule, in particular, relates to the intention of a modeler and, therefore, is specific to the notion of a model being *perspicuous*. Fig. 4 illustrates the transformations for exemplary cases.

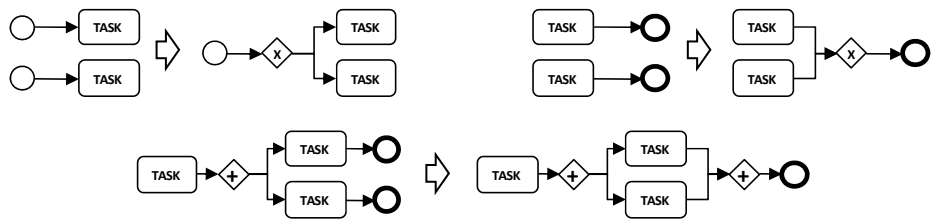

**Fig. 4.** Transformations related to the handling of start and event events.

---

[6] Note that these transformation rules may be generalized to any kind of modeling language.

**Split and join semantics.** BPMN allows for modeling nodes with more than one incoming or outgoing flow. To translate the BPMN model into a WF-net, we make those split and join semantics explicit:

- *Transform multiple incoming (outgoing) flows to an event or activity into one incoming (outgoing) flow, by preceding (following) the corresponding object with an XOR-join (AND-split) gateway that has all the incoming (outgoing) flows of the object.* [24]
- *If we determine only one origin (destination) for the multiple incoming (outgoing) flows, we use the sign of this gateway.*

Again, the latter transformation relates to the modeler's intentions. We deviate from the standard processing, if the model structure provides a strong hint to do so. Fig. 5 illustrates the transformations. In the example in the lower half, none of the split gateways qualifies to induce the type of the join gateway, so that the default transformation applies.

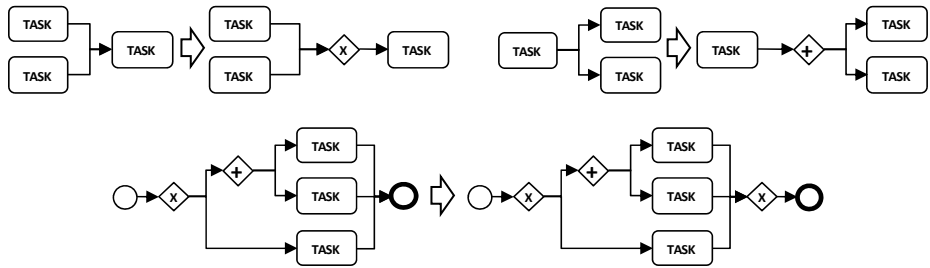

**Fig. 5.** Transformations related to split and join semantics.

**Mixed gateways.** BPMN allows for the specification of mixed gateways that combine split and join semantics. Those may be split up into a pair of a join and a split gateway of equal type [24]. However, we do not adopt this transformation for several reasons. When building the conjectures based on preliminary studies, we observed that modelers would often be unsure about semantics of mixed gateways. In contrast to the handling of start and end events and split and join semantics mentioned earlier, however, the process model structure does not provide a strong hint on the modeler's intentions regarding a mixed gateway. As such, mixed gateways lead to a non-perspicuous model. Note that those considerations are in line with the recommendation of the BPMN specification not to use mixed gateways ([27], p288).

## 4     Experimental Results

In this section we summarize the results of our experiment. Section 4.1 describes the experiment. Section 4.2 presents the results, while Section 4.3 provides a discussion.

### 4.1   Modeling Session in Eindhoven

In order to test our conjectures, we designed an experiment that would rely on the use of Cheetah Experimental Platform. The task in this experiment was to create a formal process model in BPMN from an informal description. The object that was to be modeled was the process of preparing the take-off of an aircraft[7]. We decided to use a subset of BPMN for our experiment and provided no sophisticated tool features (e.g. automated layout support or automatic syntax checkers) to prevent the modelers to become confused or overwhelmed with tool aspects [14]. A pre-test was conducted at the University of Innsbruck to ensure the usability of the tool and the understandability of the task description. This led to some minor improvements of Cheetah Experiment Platform and a few updates to the task description.

The modeling session was conducted in November 2010 with 103 students following a graduate course on Business Process Management at Eindhoven University of Technology. The modeling session started with a modeling tool tutorial, which explained the basic features of the platform. After that, the actual modeling task was presented according to which the students had to model the process shown in Fig. 2. By conducting the experiment during class and closely monitoring the students, we mitigated the risk of external distractions that might otherwise have affected the modeling process. No time restrictions were imposed on the students.

### 4.2   Results

We used the collected data of the experiment to calculate the values of the six process-based metrics of Section 3.2 for the modeling process of each student. We also determined for each modeling process the value (0 or 1) for the perspicuity metric as a measurement of process model quality. As it turned out, 54 students (52%) managed to create a perspicuous model while the remaining 49 (48%) did not.

As a next step, we looked at the distribution of the metrical values. All distributions deviated from normality, being more skewed than a characteristic Bell-curve. Therefore, we turned to the representation of these distributions as boxplots [28]. A *boxplot* (a.k.a. a *box and whisker plot*) consist of a *box*, which represents the middle 50 percent of the data. The upper boundary (also known as the *hinge*) of the box locates the 75th percentile of the data set, while the lower boundary indicates the 25th percentile. The area between these two boundaries is known as the *inter-quartile range* and this gives a useful indication of the spread of the middle 50 percent of the data. There is also a line in the box that indicates the *median* of the data (which may coincide with a box boundary) and a cross that indicates the *average value*. The *whiskers* of the box-plot are the horizontal lines that extend from the box. These indicate the minimum and maximum values in the dataset. If there are *outliers* in the data, shown as open rectangles, the whiskers extend to their maximum of 1.5 times the inter-quartile range. The boxplots for all metrics are shown in Fig. 6, 7 and 8.

---

[7] The case description is available at: http://bpm.q-e.at/experiment/Pre-Flight.

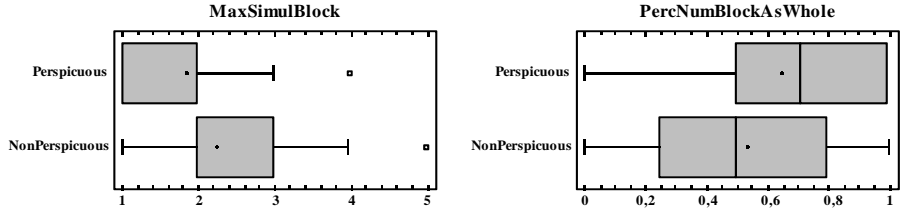

**Fig. 6.** Boxplots of the metrics for conjecture 1

What can be seen in Fig. 6 is that people who created perspicuous models tend to simultaneously work on a *smaller* number of blocks (*MaxSimulBlock*) than people who delivered a non-perspicuous model. Overall, those who developed a perspicuous model tend to complete a *higher* percentage of blocks as a whole too (*PercNumBlocksAsWhole*). Both aspects provide support to conjecture 1.

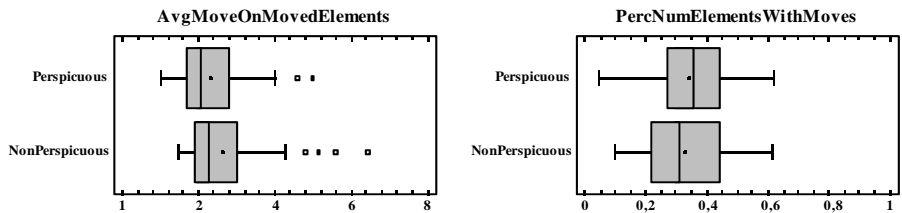

**Fig. 7.** Boxplots of the metrics for conjecture 2

In Fig. 7 it can be seen that modelers of perspicuous models tend to *less frequently* move elements than the other modelers (*AvgMoveOnMovedElements*); this is in line with conjecture 2. The groups, however, do not seem to differ very much with respect to the overall *number* of elements being moved around (*PercNumElementsWithMoves*). This can be seen from the distributions that cover about the same area. So, this gives no additional support for conjecture 2.

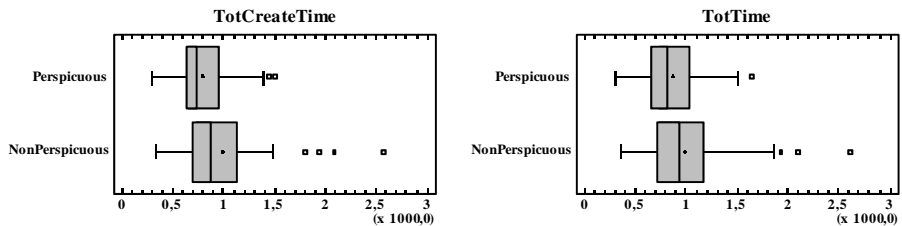

**Fig. 8.** Boxplots of the metrics for conjecture 3

Finally, Fig. 8 shows that the total time between the first and last recorded action of the modeling process (*TotTime*), as well as the total time between the first and last recorded *create* action of the modeling process (*TotCreateTime*), seem slightly lower for the group of modelers who created perspicuous models. It is this insight, i.e., that both distributions for modelers of perspicuous models cover a relatively lower range, that supports conjecture 3.

While these visual insights are promising, it is necessary to subject these to more rigorous testing. For this purpose, we carried out a t-test[8] for each of the six metrics in order to compare the respondents who created a perspicuous model with those who delivered a non-perspicuous model. The results are shown in Table 2.

What can be derived from these results is that there is a significant difference between the groups for all investigated metrics when assuming a 95% confidence interval (i.e., the P-values are lower than 0.05), except for *PercNumElementsWithMoves* (P-value equals 0.648 >> 0.05). In other words, the group of modelers who created a perspicuous model scored *significantly different* than the group who delivered non-perspicuous models with respect to *all our measures but one*, and in *exactly the direction* we conjectured. For example, the respondents who created a perspicuous model indeed were working on a lower maximum number of blocks simultaneously (*MaxSimulBlock*) and completed more blocks as one related whole (*PercNumBlockAsAWhole*) than the other group. From these results, we conclude that we have found strong support for conjectures 1 and 3 (i.e., through support for all related metrics), and mild support for conjecture 2 (i.e., via support for just one of the two related metrics).

**Table 2.** Results student t-test.

| Conjecture | Metric | T-value | df | P-value (sig.) |
|---|---|---|---|---|
| C1 | MaxSimulBlock | -2.231 | 101 | 0.028[*] |
|  | PercNumBlockAsAWhole | 2.199 | 101 | 0.030[*] |
| C2 | AvgMoveOnMovedElements | -1.984 | 101 | 0.049[*] |
|  | PercNumElementsWithMoves | 0.457 | 101 | 0.648 |
| C3 | TotTime | -2.183 | 101 | 0.031[*] |
|  | TotCreateTime | -2.505 | 101 | 0.014[*] |

(*) statistically significant values at the 95% confidence level

### 4.3 Discussion

Our findings warrant a reflection on their potential impact on research and practice. From a scientific point of view, our study confirms that the properties of a modeling process can be related to its outcome. Specifically, our work shows that aspects of a modeler's style can be operationalized and quantified, providing means to distinguish between more and less effective approaches to create a process model. As such, this work opens the venue towards a more sophisticated understanding of what makes someone a good modeler or, more precisely, what is a good modeling process. Values, beliefs, cognitive abilities, and personality traits may be as important in the field of process modeling as they are in the area of computer programming (see [29]).

---

[8] In large samples, the t-test is valid for any distribution of outcomes [32], even if we can not assume normality as is the case here.

It is also noteworthy that the attractive aspect of structured modeling in particular echoes the large interest for the formal property of *structuredness* in the process modeling field [30, 31].

From a practical point of view, our findings suggest, cf. the support for conjecture 1, that an approach that emphasizes successive phases of thorough and localized modeling (i.e., within blocks) is more attractive than diverting one's attention across different parts of a model at the same time. Similarly, yet less pronounced via mild support for conjecture 2, excessive reshaping of a model and moving its elements around seem to be anathema to good modeling practice. These are both actionable items that can be shaped into modeling instructions, which can be incorporated in process modeling courses (beyond the more traditional syntactical and formal topics). Our insight with respect to modeling speed, cf. the support for conjecture 3, seems particularly relevant to distinguish more from less proficient modelers. Such an insight may be particularly useful when composing project teams (a fast modeler is an asset, both time- and quality-wise) or assigning modeling tasks to professionals (a faster modeler will deliver a readable model).

The interpretation of our findings is presented with the explicit acknowledgement of a number of limitations to our study. First of all, our respondents represented a rather homogeneous and inexperienced group. Although relative differences in experience were notable, the group is not representative for the modeling community at large. At this stage, in particular, the question can be raised whether experienced modelers follow a similar approach to process modeling as that of skillful yet inexperienced modelers. Note that we are cautiously optimistic about the usefulness of the presented insights on the basis of modeling behavior of graduate students, since we have established in previous work that such subjects perform comparably in process modeling tasks as some professional modelers [7].

We cannot claim construct validity: In our approach we derive process metrics at the syntactical level of recorded actions of a modeler and we needed to make slight assumptions on the modelers' intentions to calculate our metrics. Nevertheless, we are hopeful that we can verify the results in later experiments, because the t-tests provided significant results (except for *PercNumElementsWithMoves*).

## 5   Conclusion

This paper reports on research about the *process of process modeling* by examining relations between the modeling process and the modeling outcome (i.e., a process model). We have been particularly interested in the notion of understandability as a quality criterion for process models and searched for related properties of the modeling process that would ensure an *understandable* modeling result.

We formulated three conjectures, i.e., that (i) structured modeling ties to model quality, whereas (ii) lots of movement of modeling objects, and (iii) low modeling speed relate to low model quality. To validate or reject these conjectures, we performed an experiment with 103 modelers and recorded for each modeler all the actions performed with the modeling tool. This allowed us to measure the related concepts of our conjectures (i.e., structuredness, movement, speed, and understandability) in metrics on the modeling process the modeling result. T-tests point at significant differences, in line with our conjectures about the quality of the model in terms of its perspicuity. We believe this provides firm empirical support for two of our conjectures and, to a lesser extent, for the remaining one.

This paper forms a basis for a deeper understanding of the process of process modeling and its impact on the quality (*in casu* understandability) of the resulting process model. If we manage to better comprehend the factors that directly influence the result of the modeling process, we would be able to comprise this knowledge in training and tools supporting process modeling. This, in turn, could result in more understandable process models, as well as a more efficient modeling process.

In this paper, we have limited ourselves to visual inspection of the distributions and t-tests to study three conjectures. Future work will include additional statistical tests on the collected data set to identify further factors describing the process of process modeling and to assess their influence on the quality of the resulting process model. Next to a further investigation of the collected data set, we will focus on validating our observations in modeling sessions while varying the modeling task to be able to generalize our findings. We also wish to include modeling experts to be able to observe a more heterogeneous group of modelers during the act of modeling.

What is also open to further study is how effective modeling instructions can be developed on the basis of our findings. Beyond instruction, we expect that tool support may be another important ingredient in achieving good modeling practice.

## 6      Acknowledgements

We wish to thank Jan Recker for his advice with respect to the development of the research method, as well as all the modelers for their contribution to this experiment.